\documentclass[a4paper,12pt]{article}
\usepackage{temp}
\usepackage{multirow}
\begin{document}

\title[Towards Threshold Key Exchange Protocols] 
      {Towards Threshold Key Exchange Protocols}
      {Towards Threshold Key Exchange Protocols}
      {Towards Threshold Key Exchange Protocols}

\author{D.\,N.~Kolegov, Y.\,R.~Khalniyazova, D.\,A.~Varlakov}
       {Kolegov~D.\,N., Khalniyazova~Y.\,R., Varlakov~D.\,A.}
       {Kolegov~D.\,N., Khalniyazova~Y.\,R., Varlakov~D.\,A.}
       
\email{dnkolegov@gmail.com, yulia.khalniyazova@gmail.com, dsurv@yandex.ru}
\udc{XXX.XX \protect\hfill 
DOI 10.17223/20710410/XX/\thePaperNo} 
\organization{BI.ZONE, Tomsk State University, Tomsk, Russia}

\makeetitle

\enabstract{
Threshold schemes exist for many cryptographic primitives like signatures, key derivation functions, and ciphers. At the same time, practical key exchange protocols based on Diffie-Hellman (DH) or ECDSA primitives are not designed or implemented in a threshold setting.
In this paper, we implement popular key exchange protocols in a threshold manner and show that this approach can be used in practice.
First, we introduce two basic threshold DH key agreement schemes that provide enhanced security features in comparison with the classic DH primitive: dealerless distributed key generation, threshold shared key computation, and private key shares refreshing. We implemented the proposed DH schemes within the WireGuard protocol to demonstrate its effectiveness, efficiency, and usability in practice. The open question is the security of the proposed schemes and their instantiations from the elliptic curves used in key agreement protocols: NIST curves, Russian GOST curves, and Curve25519.
Second, we propose an idea of implementing TLS in a threshold setting that can be used instead of Keyless SSL/TLS technology, and provide the measurements of TLS key exchanges based on threshold ECDSA.
Even if we don't provide any formal definitions, security analysis, and mathematical proofs, we believe that the ideas and mechanisms suggested in this paper can be interesting and useful. One may consider this paper as a compilation of ideas we have towards the key exchange protocols, ways to improve them with the help of threshold cryptography, and solutions to go around some difficulties we met on our way.
The main intention of the paper is to start discussions and raise awareness of the challenges and problems arising when moving to threshold key exchange protocols.
\protect\enkeywords{threshold cryptography, threshold Diffie-Hellman, threshold ECDSA, Keyless SSL, WireGuard}}


\section{Introduction}
Using threshold cryptography, one can protect a cryptographic key by splitting it into $n$ shares which are stored separately. Any subset of $t + 1$ parties each of which holds a key share can perform a cryptographic operation using the key without ever exercising or reconstructing it. At the same time, a subset of $t$ parties cannot perform such operations or get any information on the key.

We use the advantages of the threshold cryptography to strengthen DH-based key agreement and ECDSA signature, which are widely used to establish shared secrets in cryptographic protocols such as TLS and WireGuard.

Unfortunately, as we were not able to find practical and implemented threshold DH primitives, we present two simple threshold DH schemes that we've developed. 

Naive threshold DH scheme is a simple $n$-out-of-$n$ additive secret sharing scheme. It allows sharing a key among exactly $n$ parties so that no less than $n$ parties can perform an operation using the key. The threshold DH scheme is based on Shamir's secret sharing scheme and distributed key generation from \cite{tecdsa} and can be used with an arbitrary threshold value. Both schemes consist of three protocols: a distributed key generation protocol used to generate a public key and the corresponding private shares, an exchange protocol used to perform a DH operation, and a re-sharing protocol used to update the private shares. It should be noted, that the shared private key is never reconstructed during these protocols.

\subsection{Limitations}
It is worth mentioning that no single primitive can solve all security problems present in the world, and the threshold cryptography approach is no exception to this. We discuss ways to maintain a certain level of security for long-term private keys going through key exchange protocols, but the discussion of how to achieve security for the result of the key exchange protocols, namely session keys, is beyond the scope of this paper.

\subsection{Preliminaries}
In the paper, we mostly adopt the notation from \cite{tecdsa}.

A commitment scheme allows committing to a value keeping it secret with the ability to reveal the value later. Let $Com(M)$ be a commitment algorithm with an input message $M$. It outputs a commitment string $C(M)$ and decommitment string $D(M)$ which is kept secret until opening time. We denote an output of the commitment algorithm as $[KGC, KGD]$ and $[EXC, EXD]$ when used in the keygen and exchange protocols respectively. 

\subsection{Our Contributions}
In this paper, we work towards threshold key exchanges using DH and ECDSA primitives.  We claim the following main contributions:
\begin{itemize}
    \item we present two threshold DH key agreement schemes;
    \item we consider the security of instantiating threshold DH on different elliptic curves;
    \item we introduce the idea of TLS based on threshold ECDSA that can be used to reach the same security goals as Keyless SSL.
\end{itemize}

\section{Threshold Diffie-Hellman}

\subsection{Naive Diffie-Hellman Scheme}
Naive threshold DH scheme is a trivial secret sharing scheme with threshold value $t = n - 1$. A private key is generated in a distributed dealerless way so that the summing of all key shares gives the key. One of the features of the distributed key generation approach compared to key sharing through a dealer is that the generated key depends on the input of many parties.

\subsubsection{Key Generation Protocol}
The key generation protocol is used to share a private key between parties and compute the corresponding public key. To get the public key, each party computes its own public share and sums up all of the public shares received from parties. 
Input parameters for each party are a group of points on an elliptic curve $E$ with a base point $G$ of prime order $q$.
\begin{enumerate}
    \item Each party selects $x_i \in_R \mathbb{Z}_q$; computes $X_i=x_i*G$, $[KGC_i, KGD_i]=Com(X_i)$ and broadcasts $KGC_i$.
    \item Each party broadcasts $KGD_i$. The public key is set to $X=\sum_{i} X_i$.
\end{enumerate}

\subsubsection{Exchange Protocol}
The goal of the exchange protocol is to compute a shared secret with DH.
In the same fashion as in the key generation protocol, each party computes its own part of the shared secret using its private share and sums all of the pieces to get the shared secret. Let $Y$ be the remote party's (a party we are establishing a shared secret with) public key. Each local party is provided with $Y$ at the beginning of the protocol.
\begin{enumerate}
    \item Each party computes $S_i=x_i*Y$, $[EXC_i, EXD_i]=Com(S_i)$ and broadcasts $EXC_i$ to other parties.
    \item Each party sends $EXD_i$ to other parties. The DH shared secret is set to $S=\sum_iS_i$.
\end{enumerate}

\subsubsection{Re-sharing Protocol}
Re-sharing protocol provides an ability to redistribute private shares, change the number of parties $n$ or threshold value $t$ so that the public key is not changed. There are two sets of parties involved: old committee members hold private shares at the moment, new committee members will hold new private shares after the protocol completion.

\begin{enumerate}
    \item Each old party $P_i$ generates a new set of additive shares from its private share $x_i = \sum_j z_{ij}$ and broadcasts the public key $X$.
    \item Each new party notifies the old parties of its readiness for the next stage.
    \item Each old party $P_i$ sends the private share $z_{{i}j}$ to the new party $P_j^{'}$.
    \item Each new party $P_j^{'}$ takes the sum of the shares received during the previous stage to be its private share $x_j^{'}= \sum_i z_{ij}$.
\end{enumerate}

\subsection{Threshold Diffie-Hellman Scheme}
We now give consideration to the $(t, n)$-threshold DH scheme based on the distributed key generation protocol from \cite{tecdsa}, which itself is based on Feldman's VSS scheme. Input parameters for parties are a group of points on an elliptic curve $E$ with a base point $G$ of prime order $q$.
\subsubsection{Key Generation Protocol}
\begin{enumerate}
    \item Each party $P_i$ selects $u_i \in_R \mathbb{Z}_q$; computes $[KGC_i, KGD_i] = Com(y_i)$, where $y_i = u_i*G$, and broadcast $KGC_i$.
    \item Each party $P_i$ broadcasts $KGD_i$. The party $P_i$ performs a $(t, n)$ Feldman's VSS of the value $u_i$. The public key is set to $y = \sum_i y_i$. Each party adds the private shares received during Feldman's VSS protocols. The resulting values $x_i$ are a $(t, n)$ Shamir's secret sharing of the secret key $x= \sum_i u_i$. 
\end{enumerate}
\subsubsection{Exchange Protocol}
The exchange protocol is similar to the one from the naive DH scheme.
\begin{enumerate}
    \item Each party $P_i$ computes Lagrangian coefficient $\lambda_i$ and $w_i=\lambda_i*x_i$, which maps the share into a $(t - 1, t)$ share of $x$. The party computes $S_i=w_i*Y$, $[EXC_i, EXD_i]=Com(S_i)$ and broadcasts $EXC_i$. 
    \item Each party sends $EXD_i$ to other parties. The DH shared secret is set to $S=\sum_iS_i$.
\end{enumerate}

\subsubsection{Re-sharing Protocol}
\begin{enumerate}
    \item Each old party $P_i$ performs a $(t, n)$ Feldman's VSS of the value $\lambda_i * x_i$, broadcasts verification data and the old public key $X$.
    \item Each new party notifies the old parties of their readiness for the next stage.
    \item Each old party sends the private shares to the new parties.
    \item Each new party takes the sum of the private shares received during the previous stage to be its new private share.
\end{enumerate}

\section{Instantiating Threshold Diffie-Hellman Schemes}

\subsection{Threshold Diffie-Hellman on Curve25519}
Curve25519 is an elliptic curve designed specifically for the elliptic-curve Diffie-Hellman protocol (ECDH). Due to its high security and efficient scalar multiplication algorithm, it is widely adopted in modern cryptographic protocols such as TLS 1.3 and WireGuard.

Curve25519 specification requires \cite{curve25519}
setting certain bits of the scalar before scalar multiplication operation to prevent attacks (see e.g. \ref{ssa}, \ref{time}) against ECDH. 
\lstset{float=ht,
    caption={}, label={pyprog}}
\begin{lstlisting}
private_key[0] &= 248
private_key[31] &= 127
private_key[31] |= 64
\end{lstlisting}
\begin{center}
\small Listing~1. Little-endian bit clamping
\end{center}
Although it is not an issue for the private key shares in our threshold DH schemes, one may consider the absence of bit clamping harmful for the security of the virtual party's private key.

\subsubsection{Small Subgroup Attack Considerations}
\label{ssa}
The first bit clamping operation sets to zero three of the least significant bits so that the scalar is a multiple of the curve co-factor. It is aimed to protect the scalar from a small subgroup attack which allows an attacker to know three of the least significant bits of the scalar. In the case of clamped scalars, all of these bits are equal to zero. This method is simple and fast but is not convenient for non-classic schemes. Moreover, it is arguable how critical for security properties is the disclosure of three of the least significant bits. 

The following considerations were taken into account and employed \cite{decaf, modcrypto}:
\begin{itemize}
    \item successful small-subgroup attack will either result in a point known to the attacker (if the scalar is known to be divisible by the curve cofactor), or worse it may give the attacker information about the scalar. If the scalar is a private key, then leaking a few bits is a minor problem \cite{decaf};
    \item making scalar a multiple of the cofactor isn't that important for traditional X25519 DH, since a small-subgroup attack would only leak a few low bits of the scalar \cite{modcrypto}.
\end{itemize}

In our schemes, a public key is equal to the sum of public shares (possibly multiplied by the corresponding Lagrangian coefficient), which do not leak any information on the bits of the corresponding scalar, since generated private shares are multiples of the cofactor by definition of Curve25519 secret keys.

\subsubsection{Side Channel Attacks considerations}
\label{time}
The fast scalar multiplication algorithm is based on a mechanism called the Montgomery ladder and uses only one coordinate of a point. One of the first steps of the Montgomery ladder mechanism is to find the most significant bit of the scalar. By the time the search took, an attacker can guess the position of this bit. Having the chosen bit set for all of the scalars, we make sure that the timing attack won't work.

In our threshold schemes, as in the private key is never reconstructed, and therefore can not be targeted by the timing attack. 

\subsubsection{Point Addition and Coordinate Sign}
Unlike the scalar multiplication operation, the algorithm for point addition that uses only one coordinate of the point doesn't exist. For classic DH protocols, it is not an issue, since point addition is not used in such protocols. However, the threshold DH scheme implies the usage of point addition. To address this problem, we use the corresponding Edwards curve.

It is known that for every Montgomery curve there exists Edwards twisted curve, such that points on a Montgomery curve can be converted to the points on the corresponding Edwards twisted curve and vice versa. To convert the point, we need to know both of the point coordinates, which is not possible in the case of Curve25519, since all of the existing implementations use a single coordinate scalar multiplication mechanism.

However, we can recover the second coordinate by its sign. The coordinate is said to be positive if it is multiple of two, and negative otherwise. From the curve equation, we can derive two possible values of the second coordinate. Since those values have different parity, we can choose the right coordinate if we know its sign.

\subsubsection{Baby Sharks Attack Consideration}
An adversary who controls at least one of the parties could inject a small order element into the public key or shared secret by sending a low order point instead of the actual share. To prevent that, we adopt the technique from \cite{babysharks}.

The main idea is to multiply each received group element by the cofactor and then multiply it by the inverse of the cofactor to sanitize the low order group element.

\subsection{Threshold Diffie-Hellman over the NIST curves}
We use NIST's P-256 curve as a default curve for the computations. The P-256 curve lacks a cofactor or any other specialties that could affect the schemes and hence is a handy default choice for instantiating.

\subsection{Threshold Diffie-Hellman over the GOST curves}
We have successfully tested the way to instantiate our scheme over the GOST curves by "thresholdizing" the VKO functionality which is used to establish a shared key in the GOST schemes. However, a complete implementation of the threshold DH scheme over the GOST curves stays a matter of future research.

\subsection{Broadcast channel requirements}
To be correctly implemented, the proposed threshold schemes require the presence of a reliable broadcast channel\cite{tecdsa} between parties, which means that if a party broadcasts a message, it has to be guaranteed that all parties receive the same message. The simplest way to achieve this is to use an echo broadcast channel\cite{goldwasser}, which implies that all of the parties have to broadcast a hash of the message upon receiving. If all of the hashes are equal, it is considered that the message has been successfully delivered. Note, however, that such protocol doesn't provide the identifiable abort property in the dishonest majority setting (even though the current work doesn't formally specify any settings for the proposed schemes, it might be useful to know). To have this property, one has to employ a consensus protocol (e.g. Tendermint\cite{tendermint}) to satisfy the reliable broadcast requirements. 

\section{Evaluation Platform Architecture}
To evaluate the developed threshold schemes and their use within key exchange protocols we developed an evaluation platform (EP) that supports threshold DH and ECDSA schemes.

It should be noted, that we didn't focus on fault tolerance, so the platform is not even crash fault-tolerant.

\begin{figure}[htp]
    \selectlanguage{english}
    \centering
    \includegraphics[width=10cm]{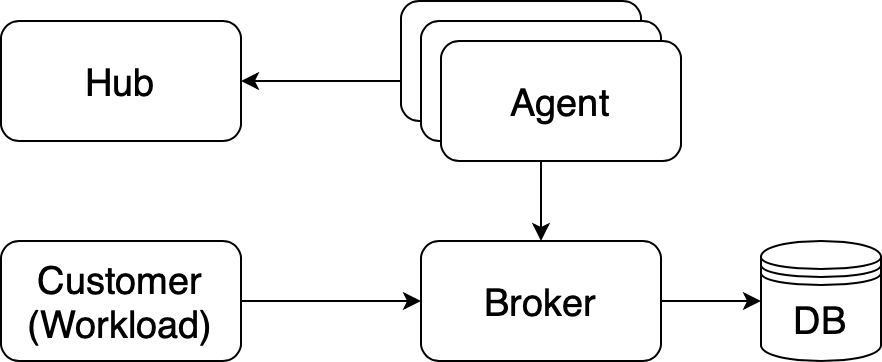}
    \caption{Components and their relation}
    \label{fig:cvsm-design}
\end{figure}

The platform consists of agents, broker and hub (see Fig. \ref{fig:cvsm-design}). Agents are parties of threshold computation following specific protocols and storing key shares. The purpose of the platform is to coordinate agents, manage access to them and provide a communication channel between them.

Broker is a component coordinating and orchestrating the agents. If a customer wants to perform a threshold computation, he connects to the broker, authenticates himself using SPIFFE and TLS, and requests the operation. Broker authenticates and authorizes requests, chooses corresponding agents to perform the computation, and retransmits the request to them. After the computation is done, agents send the result to the broker which eventually resends it to the customer.

Hub implements a broadcast channel used by agents. Hub provides private rooms, each message sent to the room is simply broadcasted to everyone in the room. After all agents have received a computation request, they also have the unique secret request ID which is used to open a dedicated room on the hub. The hub records the sent messages and sends the history to a new participant. History is erased when the computation is finished. P2P channels are emulated by encrypting messages.

The components of our platform are designed to be easily replaceable. More precisely, we wanted to have the opportunity to rewrite particular component in different languages to see how it affects performance. For this purpose, we described every protocol of interactions between components as a gRPC \cite{grpc} service in Protobuf language. It allowed us to be focused on implementing domain logic and not to care much about transport layer issues, (de)serialization, etc. Protobuf specifications can be easily compiled into the most popular languages including Go and Rust.

\section{Threshold Diffie-Hellman Evaluation}
We implemented the proposed threshold DH schemes in Go over the P-256 and Curve25519 curves, performed tests on the EP, and compared the time spent on DH operation for the classic and threshold schemes.
For measurements of the threshold schemes, we used the parameters $t = 2$ and $n = 3$ for both curves. The results of our measurements are presented in the table \ref{tab:time}.

\begin{table}[ht]
\centering
\selectlanguage{english}
\begin{tabular}{|*{5}{l|}} 
\hline
Scheme & Operation & P-256 & Curve25519 & Network parameters \\
\hline
\multirow{2}{*}{Classic DH} 
& Key generation
& $12.861\mu s$
& $100ns$
& \multirow{2}{*}{---}
\\ \cline{2-4}
& DH 
& $52.5\mu s$
& $68.942\mu s$
&
\\ \hline
\multirow{8}{*}{Naive DH}
& \multirow{4}{*}{Key generation protocol} 
& $2.258 ms$
& $21.997 ms$
& Local 
\\ \cline{3-5}
&
& $21.324 ms$
& $34. 557 ms$
& LAN 
\\ \cline{3-5}
&
& $250.288 ms$
& $154.237 ms$
& WAN 
\\ \cline{3-5}
& 
& $1.61 s$
& $1.235 s$
& Longhaul 
\\ \cline{2-5}
& \multirow{4}{*}{Exchange protocol} 
& $2.31 ms$
& $32.602 ms$
& Local 
\\ \cline{3-5}
&
& $18.151 ms$
& $44.634 ms$
& LAN 
\\ \cline{3-5}
&
& $245.41 ms$
& $165.386 ms$
& WAN 
\\ \cline{3-5}
&
& $1.604 s$ 
& $841. 412 ms$
& Longhaul 
\\ \hline

\multirow{8}{*}{Threshold DH}
& \multirow{4}{*}{Key generation protocol} 
& $3.183 ms$
& $58.15 ms$
& Local 
\\ \cline{3-5}
&
& $28.968 ms$
& $78.546 ms$
& LAN 
\\ \cline{3-5}
&
& $365.94 ms$
& $315.555 ms$
& WAN 
\\ \cline{3-5}
&
& $2.407 s$
& $2.076 s$
& Longhaul 
\\ \cline{2-5}
& \multirow{4}{*}{Exchange protocol} 
& $3.728 ms$
& $20.734 ms$
& Local 
\\ \cline{3-5}
& 
& $19.56 ms$
& $33.235 ms$
& LAN 
\\ \cline{3-5}
&
& $245.143 ms$
& $153.001 ms$
& WAN 
\\ \cline{3-5}
&
& $1.604 s$ 
& $832.781 ms$
& Longhaul 
\\ \hline
\end{tabular}
\caption{Comparison of the computation time for classic and threshold schemes}
\label{tab:time}
\end{table}

We measured the time spent on DH computations on four different sets of network parameters which are described in the table \ref{tab:params}.

\begin{table}[ht]
\centering
\selectlanguage{english}
\begin{tabular}{|*{4}{l|}} 
\hline
         & bandwidth, Mb/s  & latency, ms & MTU, bytes \\ \hline
LAN      & 100  & 2   & 1500     \\ \hline
WAN      & 20   & 30  & 1500     \\ \hline
Longhaul & 1000 & 200 & 9000     \\ \hline
\end{tabular}
\caption{Network parameters}
\label{tab:params}
\end{table}

\section{WireGuard with Threshold Diffie-Hellman}
In this section, we propose an idea of threshold WireGuard \cite{wg} that uses threshold DH operations on a long-term (static) key. The same idea can be used for TLS 1.2 or any other key exchange protocol employing DH with long-term keys. It would allow the protection of the long-term keys without the use of traditional HSM/TEE approaches.

Intuitively, threshold WireGuard is a version of the WireGuard protocol in which one or both parties use the threshold DH scheme for operations on the corresponding static private key and traditional DH for operations on the corresponding ephemeral private key.

To implement threshold WireGuard, we need a threshold version of the X25519 function. The original WireGuard protocol cannot use a threshold X25519 function directly due to the Curve25519 peculiarities: it uses only one coordinate of a point. We can go around this issue using one of the following options.

The first option is to change the WireGuard messages and semantics to send public keys in the format containing both coordinates or the first coordinate only and the sign of the second coordinate. In this case, the implementation will not be compatible with the WireGuard codebase but this is suitable for custom WireGuard protocols like Ru-WireGuard \cite{ruwg}.

The second approach is to send the sign of the second coordinate in the "reserved" field. This approach might be used if we have to be compatible with the original WireGuard protocol implementations (e.g., wireguard-go, wireguard-linux, boringtun, etc.).

The third option is to create a separate key exchange mechanism using a threshold DH scheme and emitting a shared secret established by threshold DH. This secret then will be set as a pre-shared secret (PSK) value in the original WireGuard. There are several approaches of how this can be achieved:
\begin{itemize}
    \item set up a WireGuard tunnel, run a second key exchange through it, and put the result in the WireGuard's PSK variable and perform a rehandshake;
    \item run a special threshold PSK protocol within NoiseIK pattern where peers establish a PSK using threshold DH and then will use it as the PSK for the main WireGuard.
\end{itemize}

\begin{figure}[htp]
    \selectlanguage{english}
    \centering
    \includegraphics[width=10cm]{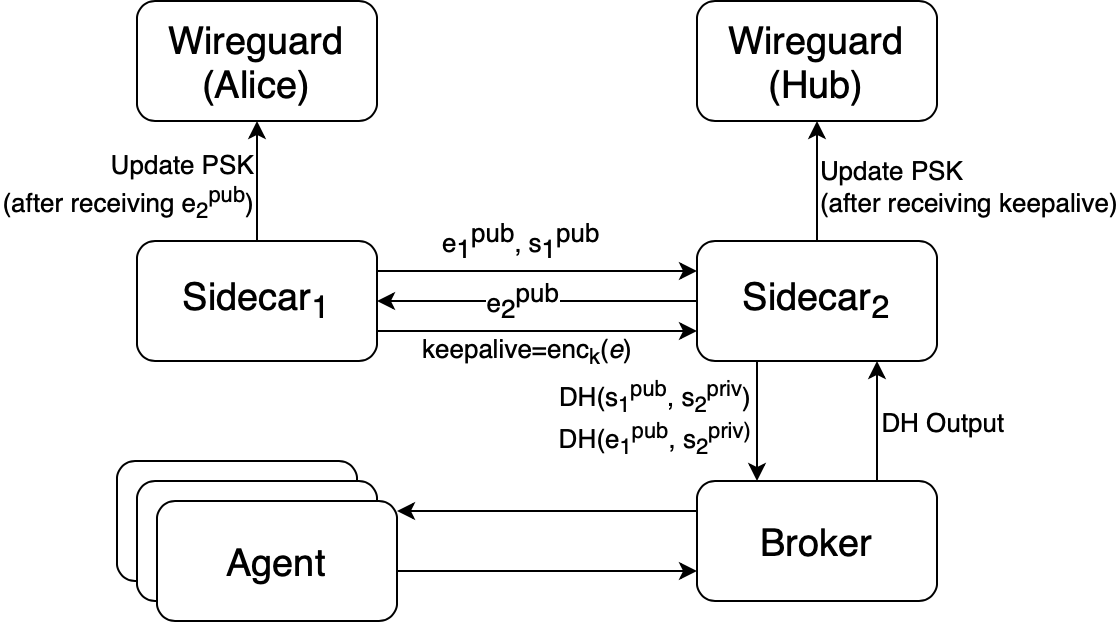}
    \caption{Running threshold PSK protocol between Alice and Hub}
    \label{fig:tptp}
\end{figure}

In this work, we've realized the latter approach of the third option (see Fig. \ref{fig:tptp}). Suppose that Alice and Bob communicate with each other via a hub using the WireGuard protocol. The hub employs a threshold DH within WireGuard to secure its static private key. The hub has the basic WireGuard keys and additional keys for threshold PSK protocol. Due to the fact that the model of the key distribution is the same (NoiseIK), Alice and Bob know the hub's WireGuard static public key. The hub's private key is shared among the parties (agents) instantiating the threshold DH scheme. When Alice connects to the hub she initiates a threshold handshake where the hub uses the threshold DH to compute a shared secret on his static private key and Alice uses the same static public key. Let PSK be the shared secret Alice and the hub established during the threshold handshake. After the PSK has been computed, Alice initiates the original WireGuard handshake with the hub using the PSK.

\section{TLS with Threshold ECDSA}
In this section, we consider how a threshold ECDSA scheme can be used in the TLS protocol to implement the Keyless SSL approach invented and widely used by Cloudflare \cite{kssl, ksslproc} to protect long-term cryptographic keys by using a threshold ECDSA scheme instead of the classic ECDSA algorithm.

The basic idea of Keyless SSL is to use a special key server controlled by a customer to perform cryptographic operations on the private key needed to perform a TLS handshake. The key server returns a message containing the result of the cryptographic operation (RSA decryption of premaster secret or ECDSA signature on randoms for DH) on the private key.

We suggest that TLS with threshold schemes can be used instead of Keyless SSL/TLS to reach the same goals.

Several threshold ECDSA schemes allow distribution of a secret key among $n$ parties and require $t+1$ of them to be cooperative to sign a digest: Gennaro-Goldfeder 18 (GG18) \cite{tecdsa}, Lindell 17 \cite{lindell17}, etc.

Consider an HTTP server working over TLS with an ECDSA-based cipher suite. Its owner would like to protect a private key and remove a single point of failure: (s)he can distribute the secret among $n$ parties. For example, a web server might be hosted on a VPS and the owner wants to be sure that the private key cannot be accessed by the provider. In this case, a private key may be distributed using the $(1, 2)$ scheme: the first party can be hosted on the same machine where the HTTP server is and the second one can be deployed on another cloud or in the customer's network. Only both parties being cooperative can perform the signing needed to finish a TLS handshake so none of the providers can access the private key.

Clearly, threshold signing will be slower as it requires several machines to do computations and talk to each other over a network, but thanks to the resumption mechanism once established, the secure channel can be reused, so for the client, it is a one-time delay.

\begin{figure}[htp]
    \selectlanguage{english}
    \centering
    \includegraphics[width=10cm]{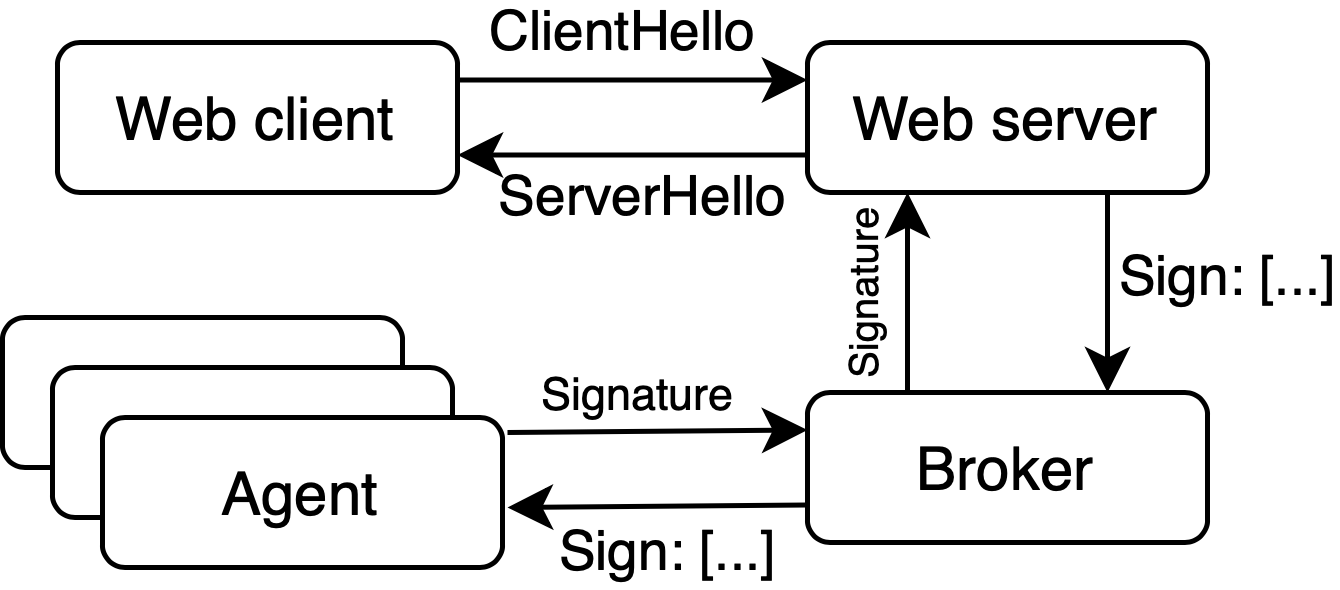}
    \caption{Deployed machines and their communications in order to perform TLS handshake}
    \label{fig:deployed-machines}
\end{figure}

We chose to use the GG18 protocol to perform threshold ECDSA as it was separately implemented and audited in Rust and Go. We implemented two kind of agents for our EP: based on tss-lib \cite{tss-lib} in Go and multi-party-ecdsa \cite{multi-party-ecdsa} in Rust. To measure their performance we deployed 6 virtual machines: 3 agents, a broker (which coordinates the agents), and a web server hosting a static page and a client (see Fig. \ref{fig:deployed-machines}). Every virtual machine had dedicated 4 CPU cores (Intel(R) Xeon(R) Gold 6142 CPU @ 2.60GHz) and 2 GB RAM.

\begin{figure}[htp]
    \selectlanguage{english}
    \centering
    \includegraphics[width=18cm]{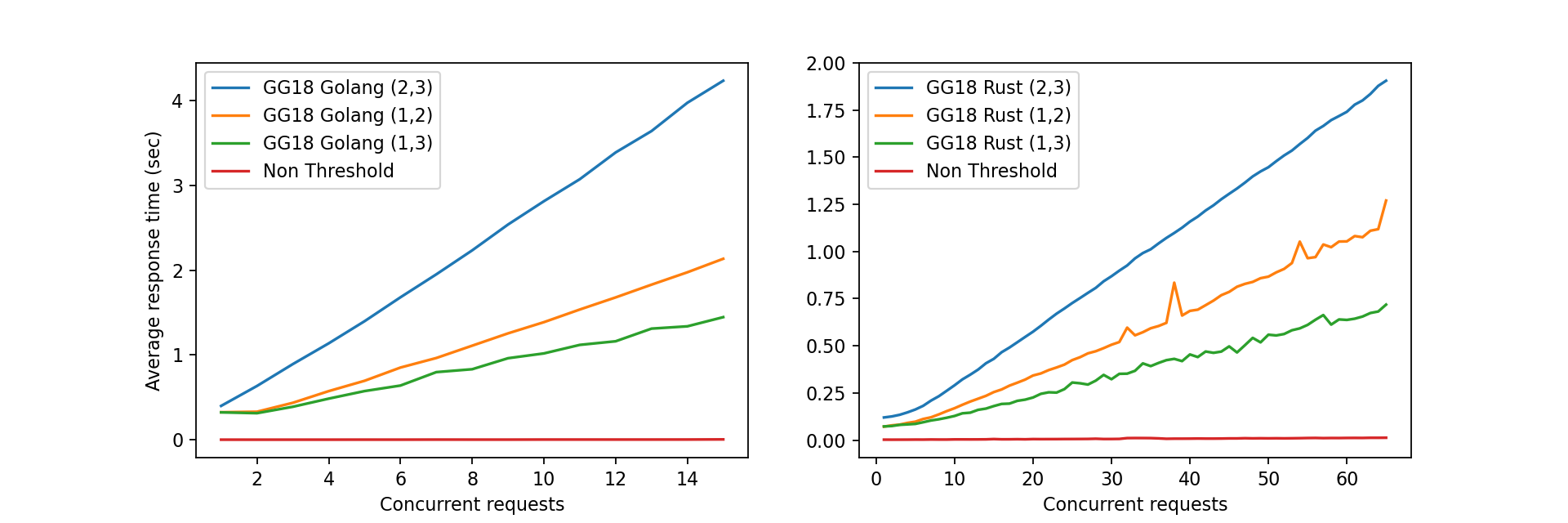}
    \caption{Web server performance in different configurations}
    \label{fig:perf-go-and-rust}
\end{figure}

\begin{figure}[htp]
    \selectlanguage{english}
    \centering
    \includegraphics[width=18cm]{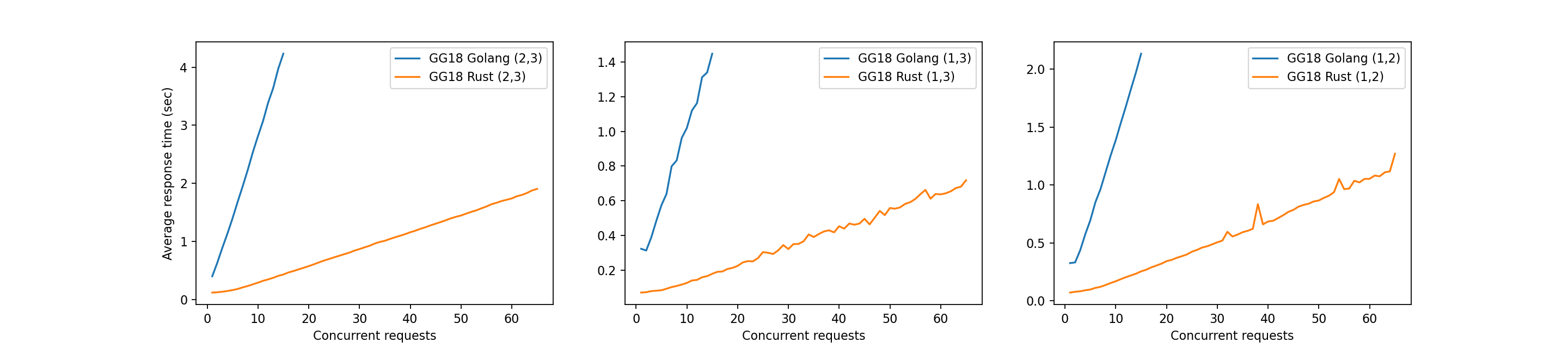}
    \caption{Comparison Rust \& Go performance}
    \label{fig:perf-go-vs-rust}
\end{figure}

We measured server average response time (from sending request to receiving response) and looked at how it changes under load by increasing number of concurrent requests. We tried different $(t, n)$ parameters: $(1, 2)$, $(1, 3)$, $(2, 3)$. The measurements are illustrated at Fig. \ref{fig:perf-go-and-rust} and Fig. \ref{fig:perf-go-vs-rust} shows the impact of using multi-paty-ecdsa comparing to tss-lib.

\section{Future Research}
The main question is the security of proposed threshold DH schemes and their instantiations on Curve25519, GOST, and NIST curves. For example, a malicious adversary could sabotage the generation of a session key by sending inconsistent public shares during the key generation and exchange protocols. As a result, the session key computed by the remote party won’t match the session key computed by the local set of parties. If this is the case, one of the possible solutions is to make all the exchange parties reveal their private shares and check the consistency of the shares. In most cases, this will allow to detection of a corrupted party and take appropriate actions. We note that approaches to handle this situation in a more efficient and reliable way stay a matter of future research.

\section{Conclusion}
There are many ways to use threshold cryptography in practice. One may split the key to prevent key leakage, apply it to achieve fault-tolerance, or enable voting functionality. We have presented two threshold DH schemes, one of which we used to instantiate WireGuard with threshold DH, and instantiated TLS with threshold ECDSA. The results and measurements show that threshold key exchange protocols can be used in practical cryptographic systems.

\begin{theenbibliography}{1}
  \bibitem{tecdsa}
    \BibAuthor{R.\;Gennaro, S\;Goldfeder.}
    \BibTitle{Fast Multiparty Threshold ECDSA with Fast Trustless Setup.}
    \BibUrl{https://eprint.iacr.org/2019/114}
  \bibitem{curve25519}
    \BibAuthor{D. J. Bernstein.}
    \BibTitle{Curve25519: new Diffie-Hellman speed records.}
    \BibUrl{https://cr.yp.to/ecdh/curve25519-20060209.pdf}
  \bibitem{feldman}
    \BibAuthor{P. Feldman.}
    \BibTitle{A Practical Scheme for Non-interactive Verifiable Secret Sharing.}
    \BibUrl{http://www.cs.umd.edu/~gasarch/TOPICS/secretsharing/feldmanVSS.pdf}
  \bibitem{babysharks}
    \BibTitle{Baby Sharks: Injecting small order points to threshold EdDSA.}
    \BibUrl{https://medium.com/zengo/baby-sharks-a3b9ceb4efe0}
  \bibitem{decaf}
    \BibTitle{Decaf: Eliminating cofactors through point compression.}
    \BibUrl{https://eprint.iacr.org/2015/673.pdf}
  \bibitem{modcrypto}
    \BibTitle{Discussion on Ed25519 "clamping" and its effect on hierarchical key derivation.}
    \BibUrl{https://www.mail-archive.com/curves@moderncrypto.org/msg00858.html} 
  \bibitem{wg}
    \BibAuthor{J. Donenfeld.}
    \BibTitle{WireGuard: Next Generation Kernel Network Tunnel.}
    \BibUrl{https://www.wireguard.com/papers/wireguard.pdf}
  \bibitem{ruwg}
    \BibTitle{Ru-WireGuard Reference Implementation with GOST Cryptographic Primitives.}
    \BibUrl{https://github.com/bi-zone/ruwireguard-go}
  \bibitem{lindell17}
    \BibAuthor{Y. Lindell.}
    \BibTitle{Fast Secure Two-Party ECDSA Signing.}
    \BibUrl{https://eprint.iacr.org/2017/552}
  \bibitem{tss-lib}
    \BibTitle{tss-lib: Golang Package Implementing Multi-Party Threshold Signature Scheme.}
    \BibUrl{https://github.com/binance-chain/tss-lib}
  \bibitem{multi-party-ecdsa}
    \BibTitle{multi-party-ecdsa: Rust Crate Implementing Multi-Party Threshold Signature Scheme.}
    \BibUrl{https://github.com/binance-chain/tss-lib}
  \bibitem{ksslproc}
    \BibAuthor{D. Stebila, N. Sullivan.}
    \BibTitle{An Analysis of TLS Handshake Proxying. 2015 IEEE Trustcom/BigDataSE/ISPA 1, 279-286}
 \bibitem{kssl}
    \BibAuthor{N. Sullivan.}
    \BibTitle{Keyless SSL: The Nitty Gritty Technical Details.}
    \BibUrl{https://blog.cloudflare.com/keyless-ssl-the-nitty-gritty-technical-details/}
 \bibitem{grpc}
    \BibTitle{gRPC: A High Performance, Open Source Universal RPC framework.}
    \BibUrl{https://grpc.io/}
\bibitem{goldwasser}
    \BibAuthor{S. Goldwasser, Y. Lindell.}
    \BibTitle{Secure Computation Without Agreement}
    \BibUrl{https://eprint.iacr.org/2002/040}
 \bibitem{tendermint}
    \BibAuthor{E. Buchman, J. Kwon, Z. Milosevic.}
    \BibTitle{Tendermint: The latest gossip on BFT consensus}
    \BibUrl{https://arxiv.org/pdf/1807.04938.pdf}    
\end{theenbibliography}

\begin{authors}
    \item{KOLEGOV Denis Nikolaevich}{PhD, Principal Research Engineer, BI.ZONE; Associate professor, Tomsk State University, Tomsk}{dnkolegov@gmail.com}
    \item{KHALNIYAZOVA Yulia Rinatovna}{Research Engineer, BI.ZONE, Tomsk}{yulia.khalniyazova@gmail.com}
    \item{VARLAKOV Denis Andreevich}{Research Engineer, BI.ZONE, Tomsk}{dsurv@yandex.ru}
\end{authors}

\end{document}